\documentclass[twocolumn,trackchanges]{aastex701}

\usepackage{graphicx}
\usepackage{xspace}

\usepackage{amsmath,mathtools}
\usepackage{xfrac}
\hypersetup{colorlinks=true,
            citecolor=blue,
            linkcolor=blue,
            urlcolor=blue
            }
\usepackage{verbatim}

\newcommand{\Msun}{$\mathrm{M}_{\odot}$\xspace}

\newcommand{\SN}{SN\,2021bnw}
\newcommand{\kms}{km\,s$^{\,-1}$}
\newcommand{\Ni}{$^{56}$Ni}
\newcommand{\Co}{$^{56}$Co}
\newcommand{\appropto}{\mathrel{\vcenter{
  \offinterlineskip\halign{\hfil$##$\cr
    \propto\cr\noalign{\kern2pt}\sim\cr\noalign{\kern-2pt}}}}}

\newcommand{\STELLA}{\textsc{STELLA}}

\shorttitle{Helium SL\SN{}}
\shortauthors{A.~Kozyreva, et al.}

\begin{document}

\title{Helium superluminous \SN{}\,: an explosion of a massive star with a pre-outburst}

\correspondingauthor{Alexandra Kozyreva}

\author[orcid=0000-0001-9598-8821,gname='Alexandra',sname='Kozyreva']{Alexandra Kozyreva}
\affiliation{Sapienza Universit\`a di Roma, Dipartimento di Fisica, 00185, Roma, Italy}
\email[show]{sasha.kozyreva@gmail.com}


\author[orcid=0000-0002-7834-0422,gname=Matteo,sname=Bugli]{Matteo Bugli}
\affiliation{Universit\'e Paris-Saclay, Université Paris Cité, CEA, CNRS, AIM, 91191, Gif-sur-Yvette, France}
\affiliation{Institut d'Astrophysique de Paris, 75014, Paris, France}
\affiliation{INAF, Osservatorio Astrofisico di Arcetri, Largo E. Fermi 5, I-50125 Firenze, Italy}
\email{matteo.bugli@iap.fr}

\author[orcid=0000-0002-9695-9835,gname=Alexey,sname=Mironov]{Alexey Mironov}
\affiliation{Sternberg Astronomical Institute, Moscow State University, 119992, Moscow, Russia}
\email{almir@physics.msu.ru}

\author[orcid=0000-0002-5920-1478,gname=Petr,sname=Baklanov]{Petr Baklanov}
\affiliation{Sternberg Astronomical Institute, Moscow State University, 119992, Moscow, Russia}
\affiliation{Astro Space Center, 117997, Moscow, Russia}
\email{baklanovp@gmail.com}




\begin{abstract}
Superluminous supernovae (SLSNe) remain an intriguing topic in supernova (SN) transient astronomy. While the majority of SLSNe are shown to be explained by energy streaming from the newly born magnetar, there are others which are powered by different mechanisms. We analyse the pseudo-bolometric light curve of  the nearby helium-rich SLSN\,2021bnw. We built models and run hydrodynamics radiative-transfer simulations with \STELLA{}. Our best-fit models include 15--22.5~\Msun{} of ejecta enriched with 1.7~\Msun{} of \Ni{} and carrying energy of 4~foe\footnote{$1\,\mathrm{foe} \equiv 10^{51}\,\mathrm{erg}$}, and colliding with 7~\Msun{} of circumstellar matter which match the observed light curve very well. The early data can be explained as cooling of an expanding shell with the mass of 0.5~\Msun{} and kinetic energy of 0.7~foe. We tend to exclude a pulsational pair-instability (PPISN) origin for SLSN\,2021bnw. Instead we conclude that SLSN\,2021bnw was preferably a core-collapse explosion of a star with the initial mass of not less than 61~\Msun{} aided by magnetorotational effects.
\end{abstract}




\section[Introduction]{Introduction} \label{sec:intro}

Superluminous supernovae (SLSNe) have become more and more popular over the last decade and remain intriguing subject in transient astronomy. By definition, their high peak luminosity exceeds absolute magnitude of $-20$~mag \citep{galyam2012}.
Mechanisms suggested to be responsible for their light curves (LCs) include primarily radioactive nickel \Ni{} and cobalt \Co{} decay, interaction between supernova (SN) ejecta and circumstellar matter (CSM), extraction of rotational energy from a central magnetar and other \citep{MoriyaSorokina2018}. 
Moreover, a number of SLSNe (together with other types of SNe) display double-peak or bumpy LCs, therefore suggesting that combinations of different mechanisms are involved in a single event. 

In the family of SLSNe there are those SNe which show identified helium lines in their spectra \citep{Yan2020}. 
Among those few He-SLSNe there is a recently discovered SN\,2021bnw \citep{Poidevin2023}, to which helium lines are clearly seen over photospheric phase \citep{Fiore2026}. 
The SN reaches a peak magnitude of $-20.5$~mags in $g$, $r$, $i$, and $o$-bands and has a broad LC spanning about 140~days.

Assuming a pure radioactive powered LC, the maximum luminosity of \SN{} of $\log L=43.6$\,[erg\,s$^{\,-1}$] requires $\sim$14~\Msun{} of radioactive nickel, which can be produced only in a pair-instability explosion of a very massive star with helium core of 110~\Msun{}.  
However, a pair-instability SN (PISN) LC is generally too slow-evolving and with a relatively high-luminosity tail, since it rises to the peak during $\sim\,100$~days and lasts more than 500~days \citep{kasenetal2011}\footnote{See also LCs for all He-core models from \citet{HW2002} in the PISN dataset \citep{dataset_PISN} 
}. 
Instead, the LC of \SN{} drops 100~days after the peak, and its tail luminosity is in fact relatively low, although there is a lack of data at late epochs.  Hence, \Ni{}\,--\,\Co{} radioactive decay cannot be the main powering mechanism for \SN{}. We note, that the magnetar spindown can mimic the \Ni{}--decay heating rate \citep{Moriya2017}, and this powering mechanism was explored for \SN{} by \citet{Poidevin2023}.

Kinetic energy coming from interaction of SN ejecta with the surrounding matter may complement \Ni{} radioactive decay energy supply (``CSM+Ni'') and amplify the luminosity of a SN. 
This combination was proposed, for instance, for SLSNe 2006gy and PTF12dam by \citet{Moriya2013} and \citet{Tolstov2017}, and other SNe \citep{Moriya2011, Noebauer2016, Leung2021}. 
Therefore, we choose to focus on CSM+Ni as possible powering mechanism for \SN{} in the present study.

In Section~\ref{sect:method} we describe the CSM$+$Ni models, and in Section~\ref{sect:results} present the resulting LCs which explain the LC of \SN{} well. 
We connect the derived ejecta and explosion characteristics to the progenitor scenario in Section~\ref{sect:evol}, which most likely indicate an explosion of a massive star via magnetorotational core collapse (MR-CC) rather than a PPISN.

\section[Light curve modeling: two main powering contributors]{Light curve modeling: two main powering contributors} 
\label{sect:method}



To build the input model for the radiative transfer simulations, we consider the SN ejecta which is unbound from the central compact object and expanding, and a CSM appended to the ejecta.

As a template for the freely expanding ejecta we use the hydrodynamical profile of a 90~\Msun{} He-model \citep{HW2002}, and construct the density-scaled profiles with different total masses. 
The hydrodynamical quantities of the ejecta are taken at day~5 after the explosion of the original model He90.  
The ejecta consist mainly of oxygen, helium, carbon, silicon, and sulfur.  
For instance, the density profile for the case of ejecta of 22.5~\Msun{} is acquired while scaling the original He90 profile by a factor of 4. 
The ejecta in this case contain: 11~\Msun{}, 3.8~\Msun{}, 3.4~\Msun{}, 2.6~\Msun{}, 1.6~\Msun{}, of O, He, C, Si, S, respectively, and 0.33~\Msun{} of \Ni{}. 
The kinetic energy, which is stored in the ejecta, is scaled with the ejecta mass accordingly, thereby corresponding to an energy of 5.9~foe. 
We do not inject any additional energy during the simulations except that produced in \Ni{} and \Co{} decay. 

We surrounded the newly constructed ejecta with a wind-like CSM ($\rho\sim r^{\,-2}$), whose inner boundary is the outer edge of the SN ejecta.  
The chemical composition of the CSM corresponds to the surface composition of the SN ejecta, namely, with the ratio of $\mathrm{He}:\mathrm{C}:\mathrm{O}=5:3:2$ (where numbers are rounded for simplicity). 
The choice of a He-rich CSM is motivated by the observed He\,I
feature in the NIR spectrum of \SN{} at day~87 after the peak \citep{Fiore2026}.


Radiative-transfer simulations for the present study are carried out with the hydrodynamics radiative-transfer code \STELLA{}  \citep{Blinnikov1998,Blinnikov2000}, which is able to follow both the radiative field and hydrodynamical evolution of the ejecta interacting with the CSM.  
The expansion of the ejecta and its interaction with the CSM obeys the standard set of non-relativistic hydrodynamic equations.  
We refer a reader to the extensive description of the code and comparison to other sophisticated radiative-transfer codes presented in \citet{Blondin2022}. 

We searched for ejecta and CSM masses suitable to reproduce the pseudo-bolometric LC of \SN{} within the ranges of 5--90~\Msun{} and 1--22.2~\Msun{}, respectively.  We also varied the mass of \Ni{} and radius of the CSM. Note that we do not carry out full parameter study.

\section{Results: SN ejecta -- CSM interaction and radioactive nickel as a working model for \SN{} LC} 
\label{sect:results}



Choosing between different combinations of ejecta-CSM, we converge to the following numbers: a CSM mass of 7~\Msun{} is required to sustain the LC around maximum phase, between $-20$~days before the peak and until 40~days after the peak;
an ejecta mass of 15--22.5~\Msun{}, which is enriched with 1.7~\Msun{} of \Ni{} needed in order to extend the LC to 100~days and support its luminosity of $\log L\sim 43.4$\,[erg\,s$^{\,-1}]$. The models with the lower-mass ejecta decline faster after day~40 and do not produce sufficient luminosity. Heavier ejecta result in too long-lasting LCs, i.e. release flux which is higher than \SN{} at this epoch. The \Ni{} mass strongly affects the luminosity during 40--90~days, i.e. the ejecta containing the amount of \Ni{} lower than 1.7~\Msun{} underestimate luminosity of \SN{} between day~40 to day~90. Certainly, \Ni{} governs the luminosity after day~100 after dropping to the tail, i.e. the ejecta containing more than 1.7~\Msun{} of \Ni{} overestimate the tail luminosity of \SN{}.

\begin{figure}[ht!]
\centering
\includegraphics[width=0.48\textwidth]{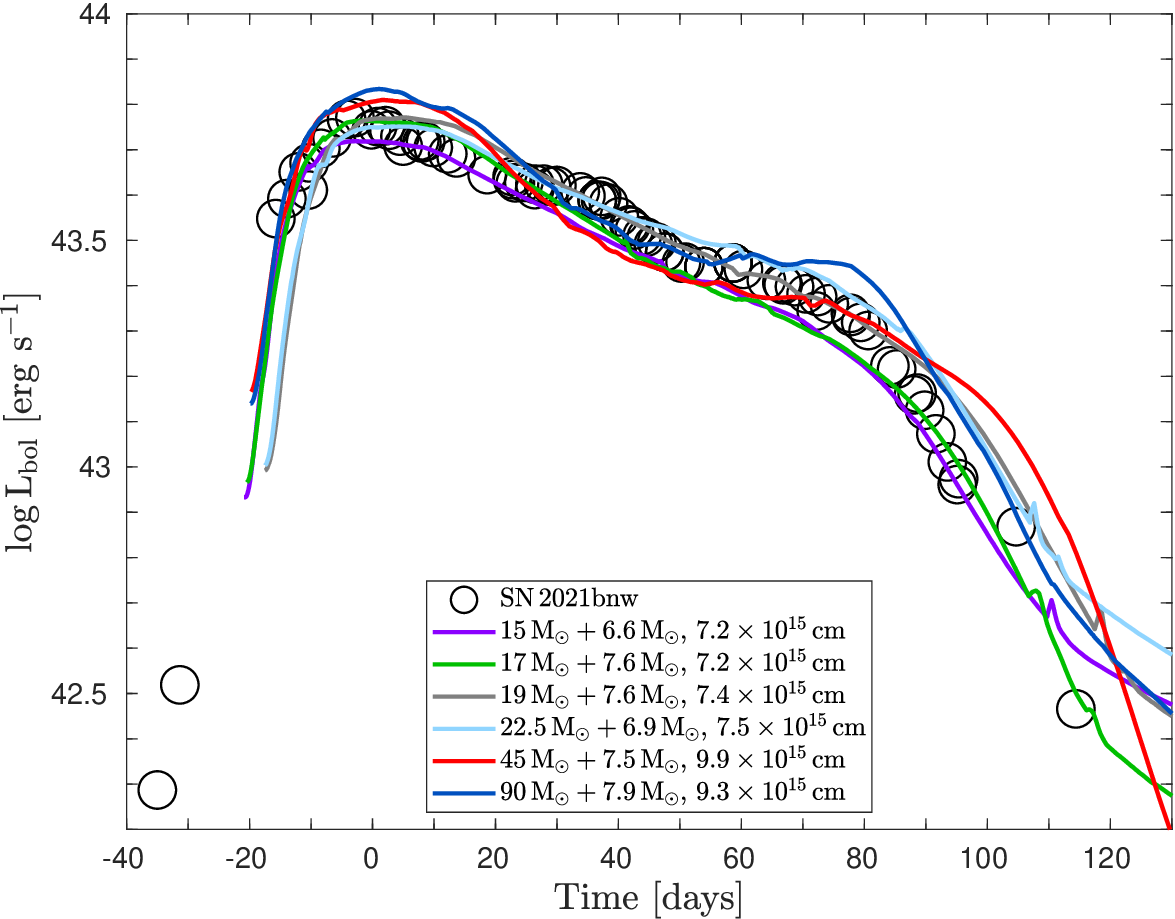}
\caption{Pseudo-bolometric LCs of the \STELLA{} models m6W1 (purple), m7W1 (green), m8W2 (grey), m2W1 (light blue), and m1W5 (red), and he90W2 (dark blue), together with \SN{} (circles). The numbers in legend are: ejecta mass, CSM mass, and CSM radius. Full list of physical parameters of the models can be found in Table~\ref{tab:stellapar}.\label{figure:main}}
\end{figure}

In Figure~\ref{figure:main}, we present the resulting pseudo-bolometric LCs of the models that match the LC of \SN{} reasonably well\footnote{All our computed models are available in the dataset \citep{dataset_2021bnw_NiCSMmodels} 
}. 
Synthetic pseudo-bolometric luminosity is integrated over the wavelength
range between 3250~\AA{} and 8900~\AA{}.
Our best-fit models m6W2, m7W1, m8W2, m2W1, and m1W5 have ejecta masses of 15~\Msun{}, 17~\Msun{}, 19~\Msun{}, and 22.5~\Msun{}, and CSM masses of $6.6-7.6$~\Msun{}. All parameters of the \STELLA{} models are summarized in Table~\ref{tab:stellapar}. We consider models which involve ejecta of 45~\Msun{} and 90~\Msun{} least probable, because unrealistic amount of energy of 12~foe and 26~foe is required to expand the ejecta. This high energy can only be produced in PISNe, which is surely not the case for \SN{}.

\begin{deluxetable}{lcccc}[ht!]
\tablecolumns{5}
\tablewidth{\textwidth} 
\tablecaption{Model parameters for \STELLA{} radiative-transfer calculations \label{tab:stellapar}}
\tablehead{
   \colhead{model} & \colhead{M$_\mathrm{ej}$ [\Msun{}]} & \colhead{M$_\mathrm{CSM}$ [\Msun{}]} & \colhead{R$_\mathrm{CSM}$ [$10^{15}$\,cm]} & \colhead{E$_\mathrm{\mathrm{kin}}$ [foe]}
   }
\startdata 
m6W1     &15   &6.6 &6.7 &  3.8 \\
m7W1     &17   &7.6 &7.2 &  4.3 \\
m8W2     &19   &7.6 &7.4 &  4.9 \\
m2W1     &22.5 &6.7 &7.3 &  5.9 \\
m1W5     &45   &7.5 &9.9 & 12   \\
W2       &90   &7   &9.3 & 26   \\  \hline
m1W3     &22.5 &6.5 &0.1 & 12   \\
m7T1     & 1   &--  &--  &  1   \\
m8T2     & 0.5 &--  &--  &  0.7 \\   
\enddata
\end{deluxetable}

To conclude, two major mechanisms are potentially required to explain the LC of \SN{}, according to our modeling and analysis: the SN ejecta--CSM interaction, and the \Ni{}\,--decay power.  A close sibling of \SN{} He-rich SNSL\,2017egm is explained as a combination of these mechanisms too \citep{Wheeler2017,Zhu2023}. \citet{Wheeler2017} found that SN\,2017egm could be explained as a
30~\Msun{} ejecta with 1~\Msun{} of \Ni{} and energy of 5~foe colliding with 0.8~\Msun{} of CSM with the outer radius of $3.5\times10^{\,14}$~cm. \citet{Zhu2023} reconsidered the modeling by \citet{Wheeler2017}, and derived a combination of the 10.7~\Msun{} ejecta containing 0.15~\Msun{} of \Ni{} and 2.7~\Msun{} of CSM.


\begin{figure}[!ht]
\centering
\includegraphics[width=0.48\textwidth]{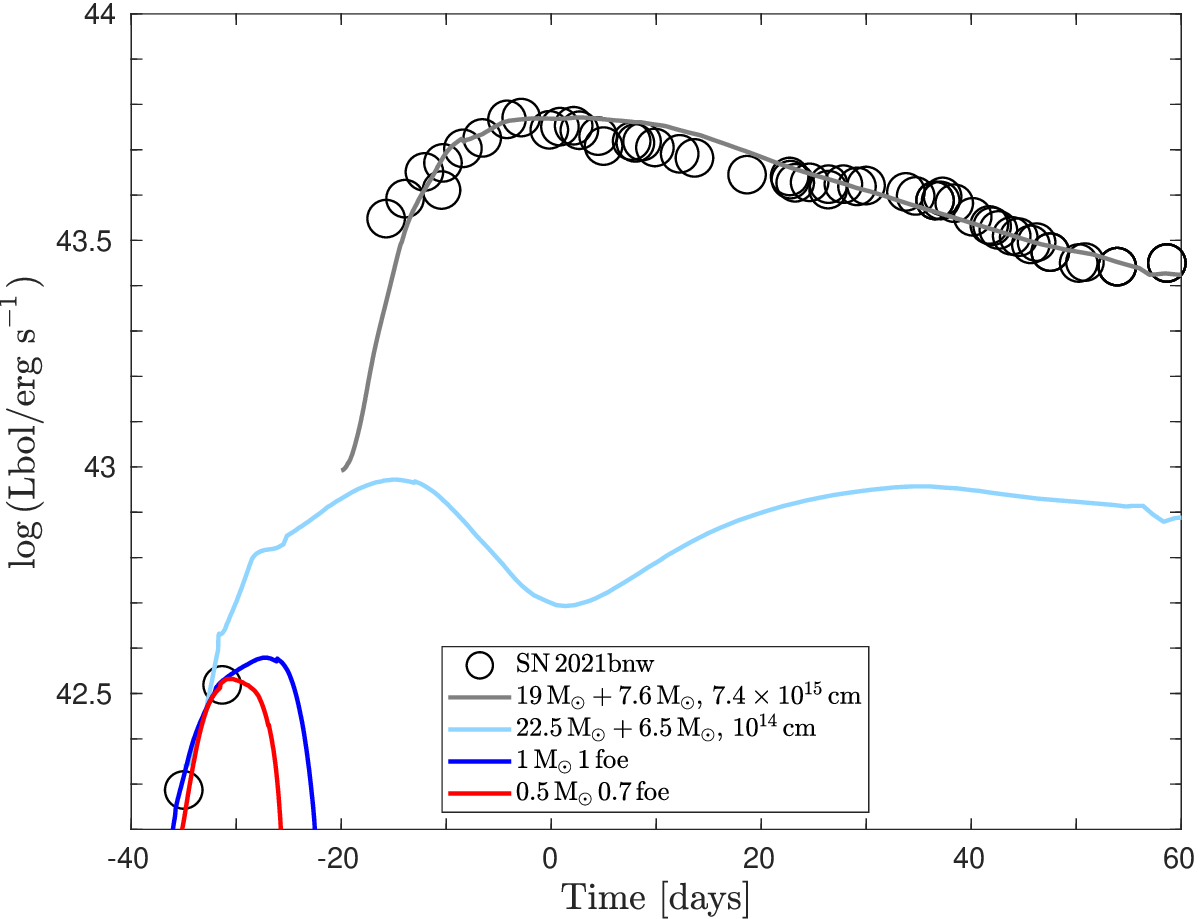}
\caption{Pseudo-bolometric LCs of \SN{} (circles) zoomed in at the early epoch, including additional models. Among models are: m1W3 (light blue) with a compact CSM, and two shell-models m7T1 (blue) and m8T2 (red). Full list of model parameters is presented in Table~\ref{tab:stellapar}. The grey curve corresponds to one of the models plotted in Figure~\ref{figure:main} (m8W2, grey).\label{figure:early}}
\end{figure}

The best-fit models discussed above reproduce the bolometric LC of \SN{} reasonably well, except for the first two epochs, about $-38$~days before the peak.  
The synthetic LCs used to explain the main LC rise from luminosity of $\log L=43$~[erg\,s$^{\,-1}$]{}, and cannot match the data points with $\log L=42.3-42.5$~[erg\,s$^{\,-1}$]{}.  
These data points are also well above luminosity of a Wolf-Rayet star ($\log L/L_\odot \sim 6 - 6.5$; see, e.g., \citealt{Crowther2007}, \citealt{Grafener2021}, \citealt{Martinet2023}), which was likely the progenitor of SN\,2021bnw.  We investigate whether a compact CSM or a pre-SN outburst might be responsible for the earlier phase of \SN{} LC.

We tested models assuming a more compact CSM with the mass of 7~\Msun{} and a radius of $7\times\,10^{\,13} - 2\times\,10^{\,14}$~cm.  We found that the model m1W3 with the CSM radius of $10^{\,14}$~cm can potentially match two early data points and show its LC in Figure~\ref{figure:early}, however, it fails to reproduce the main LC.  Optimistically, it is possible that CSM around a progenitor of \SN{} is asymmetric, and a combination of different CSM structures can explain all data of the observed LC.
However, asymmetric CSM will impact the polarimetry, which actually shows a lack of asphericity for \SN{} \citep{Poidevin2023}. Moreover, plotting the model m1W3 requires the shift of a synthetic LC by $-18$~days relative to the LCs considered as best-fit models for the main \SN{} LC. The shift of the model with compact CSM to the model with more extended CSM is nonphysical, because both models are initiated by the same explosion. Therefore, we rule out the guess about a compact CSM for the early epoch.

Taking into account the luminosity of the early data points, we explored several configurations of an expanding shell, ejected before the terminal explosion, which led to SN\,2021bnw. We varied a mass of a shell between 0.5~\Msun{} and 7~\Msun{}, 
We constructed our shell-models while scaling the density profile of the He90 model (at day~0.1 after the explosion) in the same manner as in the previous section. 
To match the observed luminosity, we varied the radius of the shell, the velocity and temperature profiles.
The resulting kinetic energy stored in the shells varied between 0.7~foe and 2.8~foe. We mapped these profiles into \STELLA{} and followed the evolution of the expanding shell and radiation field produced during its adiabatic cooling. 
The best-fit structure has an outer radius of $6\times10^{\,13}$~cm. The higher the mass of the shell, the higher the amount of energy is required to match the luminosity of the early epoch; the lower the mass of a shell, the lower the energy. 
A combination of a mass lower than 0.5~\Msun{} and corresponding energy below 0.7~foe underestimates the observed data. We choose two of our shell-models with the lowest masses of {}0.5~\Msun{} (m8T2) and 1~\Msun{} (m7T1) with the corresponding energy of 0.7~foe and 1~foe as reasonable models and show them in Figure~\ref{figure:early}.
The LCs of shell-models are shifted by $-18$~days relative to the main models.

We henceforth conclude that the first data of \SN{} can represent a pre-SN outburst, caused by an energy injection of 0.7--1~foe into an outermost layer (with the mass of 0.5--1~\Msun{}) of a progenitor about two weeks before the terminal explosion.

\section[Evolutionary path to the eruption and explosion that led to SN\,2021bnw]{Evolutionary path to the eruption and explosion that led to \SN{}} 
\label{sect:evol}


\subsection{The initial mass of a progenitor of \SN{}} 

According to our LC modelling, the \SN{} progenitor requires not less than 22--30~\Msun{} (see Table~\ref{tab:stellapar}), with an average of 26~\Msun{} H-deficient stellar {\textbf{progenitor}} before the pre-SN outburst. This estimate excludes an unknown mass of a compact object formed during the core collapse (CC). Assuming a mass of a compact object of 2.5~\Msun{}, the progenitor mass corresponds to $24.6-32.2$~\Msun{} (at the stage of helium core burning).
\citet{HW2002} derived an empirical relation between the initial zero-age main sequence (ZAMS) mass ($\mathrm{M}_\mathrm{ZAMS}$) and a corresponding He-core mass ($\mathrm{M}_\mathrm{He}$): 
$\mathrm{M}_{\mathrm{ZAMS}} = \sfrac{24}{13}\: \mathrm{M}_{\mathrm{He}}+20${}. 
This relation is valid for progenitor ZAMS masses $140-260$~\Msun{} and corresponding He-cores of $64-133$~\Msun{}, i.e. higher than 26~\Msun{} inferred from the LC modelling.  
Assuming this relation might be applicable for lower mass stellar models, the progenitor in our case would have $65-79$~\Msun{} at ZAMS
Mass-losses during the stage of H-deficient evolution should also be taken into account, however, this cannot be estimated straight-forward, since wind mass-losses are unpredictable for most massive stars \citep{Vink2015}. 
Nevertheless, if 7~\Msun{} CSM can be considered as matter lost as wind, i.e. the range for He-core is lower, and corresponding ZAMS mass is: $53-65$~\Msun{}.
Therefore, 73~\Msun{} serves as a rough estimate for the initial mass of the \SN{} progenitor.

\citet{Yusof2013} calculated a grid of very massive stellar models, and found that a star with a ZAMS mass of 120~\Msun{} at solar metallicity results in a progenitor of 30.81~\Msun{} at the end of oxygen burning, close to 26~\Msun{} related to the case of \SN{}.  The environment of \SN{} has the lower metallicity ($Z_\odot/3${}), i.e.  120~\Msun{} is an overestimated value for an initial mass of the progenitor, because of the lower expected wind mass-losses. Their 120~\Msun{} rotating model at LMC ($Z_\odot/2.3$) metallicity ended up as 39.2~\Msun{} progenitor, which is higher than needed for our best-fit models and means that the \SN{} progenitor had a ZAMS mass lower than 120~\Msun{}. Hence, we deduce that the progenitor of \SN{} was a massive star whose ZAMS mass lies above 61~\Msun{} but below 120~\Msun{}, if a single stellar evolution scenario is considered. Refining the prediction of the progenitor mass range requires detailed stellar evolution calculations, which are beyond the scope of the current study.

The terminal explosion of such progenitor should be an abnormally high (5~foe) energetic event, similar to hypernovae \citep[][]{Mazzali2006,Iwamoto1994}, during which a high mass of \Ni{} (1.7~\Msun{}) is produced, as inferred from our LC modelling. These numbers lie beyond the current limits of physically-consistent neutrino-driven core-collapse simulations \citep{Ertl2020} but are hypothetically possible, especially if magnetic fields are involved, and the core retains its angular momentum. This phenomenon is represented by a collapsar or a magnetorotational core collapse 
\citep[MR-CC,][]{Bisnovatyi1976, MacFadenyen1999, Bugli2021, Obergaulinger2021, Mueller2024}. 


The inferred mass of radioactive nickel \Ni{} required to power the LC of SN2021bnw, namely 1.7~\Msun{}, exceeds also \Ni{} yields currently obtained in physically-consistent MR-CC explosions (powered by either a proto-magnetar or a black hole): $0.5-1$~\Msun{} \citep{Grimmett2021,Reichert2023}, and up to 1.14~\Msun{} in higher-mass collapsar scenario \citep{Siegel2022}.
Nevertheless, there are several caveats, which hypothetically may lead to
higher \Ni{} masses produced in MR-CC:
\begin{enumerate}
    \item The existing MR-CC explosion simulations show the evolution within ~1 s after the bounce. However, neutrino-matter interaction continues, as a shock propagates thought the ejecta in directions different than the propagation axis of the jet, therefore, one can expect a continuous production of heavy elements during later phases, including \Ni{} \citep{Siegel2019,Obergaulinger2020Handbook}.
    \item More approximated neutrino transport approximations used in numerical models (such as leakage schemes) have been shown to overestimate the neutron-richness of the ejecta, thus obtaining higher estimates for r-process elements production at the expenses of more proton-rich elements such as \Ni{} \citep{Obergaulinger2020Handbook}.
    \item The explosion dynamics directly influences the production of heavy elements \citep[e.g.,][]{Reichert2024}.
The resolution adopted by MR-CC studies can significantly affect small-scale dynamical processes, e.g., turbulent motion and consequent efficiency of neutrino-matter interactions \citep{Varma2026}. However, 3D simulations
remain computationally challenging, as they are both limited in the resolution of their grids and in their overall duration.
\end{enumerate}
Future developments in first-principle numerical models could, therefore, lead to systematically higher \Ni{} masses produced in MR-CC simulations.

\subsection{Can \SN{} be a PPISN?} 
\label{subsect:ppisn}

The principal criterion to distinguish the PPISN and core-collapse of a massive star is the following: in He cores with the mass above 35~\Msun{}, the pressure is dominated by radiation, which allows electron-positron pairs to be produced at the onset of oxygen burning \citep{FowlerHoyle1964, Barkat1967, BisnovatyKazhdan1967, RakavyShaviv1967, Wagoner1969, BlinnikovDuninaNadezhin1996}. 
However, at this low mass limit pair creation happens in a small part of the core and little energy is generated in explosive oxygen burning 
\citep[$10^{\,48}-10^{\,49}$~erg; see Table~1 in ][]{Woosley2017}, thus leading to low ejected masses (fraction of solar mass).
Under any circumstances, there is a massive, 30--40~\Msun{} black hole left behind the terminal CC for progenitors in the PPISN range (35--60~\Msun{} He cores), without ejection of \Ni{}, and SN phenomenon as well \citep{Powell2021, Rahman2022}.
PPISN at the low-mass limit do not result in a luminous and long-lasting transient \citep{Woosley2017}.
Higher mass PPISN progenitors can have a series of high energy mass ejections, reaching 30~\Msun{} in total \citep[e.g.,][]{Marchant2019, Renzo2020}, still leaving 30--40~\Msun{} black hole at the end with no ejected \Ni{}. 
In contrast, \citet{MoriyaMarchantBlinnikov2020} consider a progenitor of 82.5~\Msun{} and suggest that their 45~\Msun{} black hole does not fully collapse,
but explodes with 10~foe and 1.5~\Msun{} \Ni{} ejected, leaving a 35~\Msun{} black hole instead.

In our modelling we consider freely expanding ejecta which carry kinetic energy, i.e. the energy defined as a residual of the amount of explosion energy and gravitational energy of the central object.
The higher the mass of a compact object (black hole), the higher should be the explosion energy to overcome the gravitational field created by the central compact object and allow the ejecta ($\sim 20$~\Msun{} in our case) to have certain kinetic energy (4--5~foe). 
Taking into account that 5~foe is already a quite high energy, we tend to conclude that a much higher energy is problematic to achieve in a realistic CC explosion.

Because we suggest Ni-power as one of the energy sources supporting the light curve of SN\,2021bnw, \Ni{} has to be produced, consequently, high-mass PPISNe are not appropriate explanation according to our scenario.
It might be that SN\,2021bnw is a combination of two episodes of interaction, therefore, PPISN mechanism can be considered suitable. We leave this possibility to explore in other studies.

Below the low-mass PPISN limit, cores evolve without significant pair production, and follow the nuclear evolution of a massive star until iron-core formation.
In our study we show that the total mass of ejecta and CSM is $\sim 26$\Msun{}.
Adding a compact object of 2.5~\Msun{} (see though Table~2 in \citealt{Aloy2021}: 3.39~\Msun{} black hole for the extreme case 35OB-RRw) results in a 28.5~\Msun{} He star before it loses 7~\Msun{} to form surrounding CSM, which is suspiciously close to the range of PPISN progenitors.
However, the core itself is about 21.5~\Msun{}, and cannot be considered as a He core of a possible PPISN progenitor.

A general consensus is that the lowest mass of a helium core which can finalise its life as a PPISN ranges between 34~\Msun{} for evolution without mass-loss, and 48.5~\Msun{} for evolution which accounts for mass-loss, rotation, and magnetic field.  We note that in any reliable PPISN scenario, the final mass of the progenitor preceding the terminal core collapse, i.e.  after a sequence of pair-instability pulses, is 33.87--43.06~\Msun{}, which entirely collapses into a black hole.  This terminal event occurs without mass ejection, ejected radioactive nickel, consequently, without a SN-like transient.

Other studies \citep[see, e.g.,][]{Yoshida2016} explore PPISN models which end up as core-collapse explosions with very high explosion energies, up to 10~foe, small black hole masses of 2~\Msun{}, final ejecta of 50~\Msun{}, and relatively large amount of \Ni{}, 1.2-2~\Msun{}. 
\citet{Tolstov2017}, \citet{Leung2019}, and \citet{MoriyaMarchantBlinnikov2020} use massive stellar models, including PPISNe, with the final progenitor mass of about 40~\Msun{}, which by default should collapse into a black hole.
Being blown up by an artificial injection of high energy (5--20~foe) and with artificially set large mass of \Ni{}, these models explain several luminous transients (PTF12dam, SN\,2011kl, AT2018cow, SN\,2018gep), in turn, interpreted as PPISNe \citep{Leung2021}. 

Our best-fit models have about 19~\Msun{} of expanding ejecta interacting with 7~\Msun{} of CSM, and we do not include a central compact object in our simulations. Since explosion calculations are beyond the scope of the study, we cannot decide what was the mass of the final pre-collapse core before the terminal explosion. Therefore, to be on the safe side, we propose that the progenitor in the case of \SN{} was well below the low-limit of the PPISN carbon-oxygen core mass of 34~\Msun{}, and it is fair to conclude that SN\,2021bnw is most likely not a PPISN \citep[see also][]{Hirschi2025}.

\subsection{The origin of CSM and the pre-SN outburst in \SN{}} 
\label{subsect:outburst}

As discussed above, we decline a possibility of \SN{} to be a PPISN, as the estimated ejecta+CSM mass lies below the low limit for a PPISN progenitor. Therefore, the progenitor is presumably a massive star which ends its life as a core-collapse SN. 
In this view the progenitor at the moment of the pre-outburst is at the stellar evolution stage of core oxygen burning, which lasts about a month and is followed by several hours of core silicon burning \citep{WHW2002,Hirschi2004}. 
The star consists mainly of a burning core with a helium-enriched outer layer which expected to be ejected.

Preceding the pre-outburst the star lost 7~\Msun{} of helium-rich atmosphere through intensive wind, which is viable for hydrogen-deficient stars \citep{Meynet2005}. 
This matter is assumed to shape the required CSM.  Further assuming a velocity for the CSM of 10--100~\kms{}\,, 20--200~yrs are needed for the CSM to spread to the radius of $7\times10^{\,15}$~cm.  This means that the CSM has been formed since core helium burning.
The origin of a possible mass ejection preceding the terminal explosion by 18~days is more speculative. There are different physical processes that could be responsible for mass ejection in massive stars. Firstly, convection driven gravity waves generate a non-linear velocity field and cause additional energy deposition into loosely bound envelope \citep{Quataert2012, Fuller2018, Matzner2021}.
Secondly, as a star has a set of nuclear burning shells, instabilities arise from turbulent convective motions coupled with shell burning \citep{Yadav2020,Varma2025, Whitehead2026}. 
Thirdly, pulsations can lead to pulsationally-driven mass-loss \citep{Fadeev2004, Yoon2010}.
These three mechanisms can be responsible, in part, for the origin of the formation of CSM as well.

\section[Conclusions]{Conclusions} 
\label{sect:conclusion}


In this study, we built descriptive models which explain the pseudo-bolometric light curve of He--SLSN\,2021bnw well. Our models combine interaction of the SN ejecta with the surrounding matter, and radioactive nickel \Ni{} power supply. $15-22.5$~\Msun{} of ejecta with kinetic energy of 4~foe smash into 7~\Msun{} of helium-rich CSM and produce the amount of energy needed to power the first 80~days of the observed LC. To support the rest of \SN{} LC until day~140, the ejecta should be enriched with 1.7~\Msun{} of \Ni{}. The early data points in the observed LC cannot be explained by interaction of the ejecta which carry such high energy, and might represent the pre-outburst which occurred 18~days before the terminal explosion, i.e. during core oxygen burning. The expanding shell with the mass of half of solar and kinetic energy of 0.7~foe emits enough energy while cooling to match these data.

The sum of the ejecta and the CSM in our best-fit models ranges between 22~\Msun{} and 30~\Msun{}. These numbers lie below the PPISN mass limit, therefore, we lean against a pulsational pair-instability origin for SLSN\,2021bnw.  Instead, we suggest that \SN{} was preferably a core-collapse explosion of a massive star with initial mass not less than 61~\Msun{} that was driven by the magnetorotaitonal mechanism (rather than neutrino heating) and produced a large amount of \Ni{}.

Current nucleosynthesis calculations from state-of-the-art extreme core-collapse SN simulations demonstrate indeed their potential in producing high amounts of \Ni{} \citep[see][for a recent review]{Obergaulinger2020Handbook}. There are, however, a few caveats that need to be considered. While the magnetorotational core-collapse SN scenario appears to be the most likely to have occurred in the case of SLSN\,2021bnw, it remains still unclear whether the explosion dynamics of a high-mass progenitor such as the one required in our model could deliver the inferred 1.7~\Msun{} of \Ni{}.
Moreover, to achieve such high yields the intense magnetic field that extracts the central engine's rotation would most likely need to be a large-scale dipolar field aligned with the rotational axis, since this appears to be the only configuration that ensures a sufficiently high efficiency in the explosion dynamics \citep[][]{Bugli2020, Bugli2023} and nucleosynthesis \citep[][]{Reichert2023, Reichert2024}.  
Future work on explosive nucleosynthesis from magnetorotational core-collapse SNe could address this scenario more specifically and therefore put to the test the results of the current study.

\begin{acknowledgments}
We are grateful to the referee for helping us to make the message of the manuscript to be clearer.
We thank Achille Fiore for providing the pseudo-bolometric LC of \SN{}. 
We thank Matt Nicholl, Steve Schulze and Thomas Janka for fruitful discussion, which helped us to improve the models and interpretation. 
This project has received funding from the European Research Council under the European Union's Horizon Europe research and innovation program (grant agreements No.~101117510 and 101164144). 
MB acknowledges the support of the French Agence Nationale de la Recherche (ANR), under grant ANR-24-ERCS-0006.
PB is supported by the grant RSF~24-12-00141 for modeling supernova light curves with the \STELLA{} code.
\end{acknowledgments}

\bibliography{references}{}
\bibliographystyle{aasjournalv7}

\end{document}